\documentstyle[12pt]{article}
\newcommand{\be}{\begin{equation}}
\newcommand{\ee}{\end{equation}}

\date{}
\begin{document}
\begin{titlepage}
\begin{flushright}
DESY 94-081\\
HD--THEP--94--6\\
hep-ph/9405300
\end{flushright}
\quad\\
\vspace{1.8cm}
\begin{center}
{\bf\LARGE SEARCH FOR THE QCD GROUND STATE}\\
\vspace{1cm}
M. Reuter\\
\bigskip
DESY, Notkestra\ss e 85, D-22603 Hamburg \\
\vspace{.5cm}
C. Wetterich\\
\bigskip
Institut  f\"ur Theoretische Physik\\
Universit\"at Heidelberg\\
Philosophenweg 16, D-69120 Heidelberg\\
\vspace{1cm}
{\bf Abstract}
\end{center}

Within the Euclidean effective action approach we propose criteria
for the
ground state of QCD. Despite a nonvanishing field strength the ground
state
should be invariant with respect to modified Poincar\'e
transformations consisting of a combination of translations and
rotations with suitable
gauge transformations. We have found candidate states for QCD with
four or more
colours. The formation of gluon condensates shows similarities with
the
Higgs phenomenon.
\end{titlepage}
\newpage

1. It has been known for a long time that the perturbative
vacuum cannot be the true ground state
of QCD \cite{1}. If one considers the Euclidean effective action
$\Gamma$ as a function of the colour-magnetic and colour-electric
fields $B^z_i, E^z_i$, one finds states with $B,E$ different from
zero for
which the value of $\Gamma$ is lower than for $B=E=0$.
This observation constitutes the basis for many models of
condensates of composite operators as $F_z^{\mu\nu}F^z_{\mu\nu}$
\cite{2}. Since the existence of states with lower energy than
the perturbative vacuum is clearly visible in the effective
action for the gluon field $A_\mu$, one may wonder what is
the true QCD ground state in this language.\footnote{In the present
context
``ground state'' means the classical field configuration which
constitutes
the absolute minimum of the Euclidean effective action. It should not
be
confused with the quantum mechanical vacuum state.}
 The Euclidean effective
action for a pure Yang-Mills theory should be bounded below and we
know
that the state $A_\mu=0$ (or a gauge-equivalent state) is not
the state of lowest action. There must therefore exist an absolute
minimum of $\Gamma$ with $F_{\mu\nu}^z\not=0$. An immediate
worry is then the apparent breakdown of Euclidean rotation
symmetry (corresponding to Lorentz-symmetry) for any nonvanishing
value of the antisymmetric tensor field $F_{\mu\nu}$. This
seems to be in contradiction with the observed Poincar\'e symmetry
(global $d$-dimensional rotations and translations in the
Euclidean language) of our world which requires a ground state
invariant with respect to these symmetries. Furthermore, no parity
$P$, time reversal $T$ or charge conjugation $C$ violation is
observed in strong interactions and the QCD ground state must respect
these discrete symmetries as well.

The situation is less dramatic if one realizes that the standard
implementation of rotations and translations is not the only
way to realize the Poincar\'e symmetry. One may define modified
Poincar\'e transformations by combining the standard space
transformations with appropriate gauge transformations. An example
for the $SO(3)$ rotation group is well known for instantons
\cite{3} where space rotations are combined with $SU(2)$ gauge
rotations to form a new rotation group. With respect to the combined
symmetry transformations the
instanton is invariant. From the observational point of view there is
no way of distinction between the new ``combined'' rotations and
the standard rotations. Furthermore, the combined rotations
act as standard rotations on any gauge-invariant state. Another
example for a ground state with a modified translation group are
spin waves \cite{4}.\footnote{Also in the classic example of the
``magnetic
translation group'' \cite{M} the generators are similarly modified.}

 The aim of the present letter is to search
for a ground state for QCD with $F^z_{\mu\nu}\not=0$ which
nevertheless preserves a new version of Poincar\'e symmetry as well
as $P, T$ and $C$ symmetry.

The four-dimensional rotation symmetry $SO(4)$ is locally equivalent
to the direct product $SU(2)_L\times SU(2)_R$ with generator
$\vec\tau_L, \vec\tau_R$ obeying
\begin{eqnarray}\label{1}
\lbrack\tau_L^i,\tau_L^j\rbrack&=&i\epsilon^{ij}_{\ \
k}\tau^k_L\nonumber\\
\lbrack\tau_R^i,\tau_R^j\rbrack&=&i\epsilon^{ij}_{\ \
k}\tau^k_R\nonumber\\
\lbrack\tau^i_L,\tau^j_R\rbrack&=&0\end{eqnarray}
It is obvious that this group structure remains unchanged
if we combine $SU(2)_L$ or $SU(2)_R$ or both with appropriate
$SU(2)$ subgroups of the gauge group $SU(N)_C$. (We discuss
here general $N$ and will later discuss special properties for
$N=3$.)
Denoting by $\vec\tau_1,\vec\tau_2$ the generators of two
groups $SU(2)_1, SU(2)_2$ commuting with $SU(2)_L,SU(2)_R$ one
obtains new rotation symmetries
\begin{eqnarray}\label{2}
SU(2)_L'&=&diag(SU(2)_L,SU(2)_{1})\nonumber\\
SU(2)_R'&=&diag(SU(2)_R,SU(2)_{2})\end{eqnarray}
with generators
\begin{eqnarray}\label{3}
\vec\tau_L'&=&\vec\tau_L+\epsilon_L\vec\tau_1\nonumber\\
\vec\tau_R'&=&\vec\tau_R+\epsilon_R\vec\tau_2
\end{eqnarray}
We will also consider the possibility that only $SU(2)_L$ (or
$SU(2)_R$)
is modified and therefore admit
$\epsilon_L,\epsilon_R=0$ or 1. The new generators $\vec\tau_L'$
and $\vec\tau_R'$ fulfil the same commutation relations
(\ref{1}) as $\vec\tau_L$ and $\vec\tau_R$ provided the subgroups
$SU(2)_1$ and $SU(2)_2$ commute
\be\label{4}
\lbrack\tau_1^i,\tau_2^j\rbrack=0\ee
\bigskip

2. Let us first investigate the
possibility that $SU(2)_{1,2}$
are subgroups of global $SU(N)_C$ transformations. Then the
group structure of $SU(N)$ implies that we can modify both $SU(2)_L$
and $SU(2)_R$ $(\epsilon_L=\epsilon_R=1)$ only for $N\geq4$. For
$N=2,3$ two commuting $SU(2)$ subgroups do not exist within the
global $SU(N)$
transformations and either $\epsilon_L$ or $\epsilon_R$ must vanish.
(We will choose $\epsilon_L=1,\epsilon_R=0$, but the
opposite choice is equivalent.) The ground-state structure
may therefore be different for $N\geq4$ and for $N=2,3$.

We begin with the case $N=4$ where $SU(4)_C$ has
a $SO(4)_C$ subgroup. (This discussion can be generalized to all
gauge
groups containing an $SO(4)$ subgroup.) The  15-dimensional adjoint
representation of $SU(4)_C$ transforms with respect to
$SO(4)_C=SU(2)_{C1}\times SU(2)_{C2}$ as

%\newpage
\begin{displaymath}
15\quad\to\quad(3,1)+(1,3)+(3,3)\qquad\qquad\qquad\qquad(I)\end{displaymath}
or
\be\label{5}
15\quad\to\quad(1,1)+(3,1)+(1,3)+(2,2)+(2,2)\qquad(II)\ee
(There are two inequivalent embeddings I, II of $SO(4)$ in
$SU(4)$.) For the embedding II one may put all gauge fields $A_\mu^z$
to zero except for those in one (2,2) representation which we denote
in the standard $SO(4)$ vector notation by $A^\alpha_\mu,\alpha=
1...4$. The state
\be\label{6}
\langle A^\alpha_\mu\rangle=a\delta^\alpha_\mu\ee
with constant nonvanishing $a$ is manifestly invariant under standard
translations as well as under the combined $SO(4)$ rotation group (2)
(with $\epsilon_L=\epsilon_R=1$).
An arbitrary space rotation acting on the index $\mu$ of $A^\alpha
_\mu$ can be compensated by an appropriate $SO(4)_C$ gauge
rotation acting on $\alpha$. (In a more group-theoretical language
$A^\alpha_\mu$ belongs to the representation (2,2,2,2)
with respect to $SU(2)_L\times SU(2)_R\times SU(2)_{C1}\times
SU(2)_{C2}$ and this representation contains a singlet (proportional
to (\ref{6})) with respect to the subgroup $SU(2)_L'
\times SU(2)_R', SU(2)_L'=diag
(SU(2)_L, SU(2)_{C1}),SU(2)_R'=diag (SU(2)_R,SU(2)_{C2}).)$
It is easy to verify that the commutation relations between modified
rotations $(\vec\tau_L',\vec\tau_R')$ and translations $P_\mu$
are the same as for the usual rotations since the gauge
transformations of $SO(4)_C$ are coordinate independent $([\vec
\tau_{1,2},P_\mu]=0)$.
The state (\ref{6}) is therefore invariant under a modified
version of Poincar\'e symmetry. (The discrete symmetries $P,T,C$ will
be
discussed later.) Hence it is a possible candidate for the ground
state of a pure $SU(4)_C$ gauge theory. The field strength for the
gauge field (\ref{6}) is easily computed with $T_\alpha$ the
generators
of $SO(4)_C$ and
\be\label{7}
A_\mu=A_\mu^\alpha T_\alpha\ee
One finds a nonvanishing value
\begin{eqnarray}\label{8}
F_{\mu\nu}&=&-ig\lbrack A_\mu,A_\nu\rbrack\nonumber\\
&=&-iga^2\delta^\alpha_\mu\delta^\beta_\nu\lbrack T_\alpha,T_\beta
\rbrack\nonumber\\
&=&ga^2f_{\mu\nu}^{\ \ \gamma}T_\gamma\end{eqnarray}
with $g$ the gauge coupling and $f_{\alpha\beta}^{\ \ \gamma}$
the structure constants of $SO(4)$. We emphasize, nevertheless,
that (\ref{6}) is not the only possible ground state candidate.
Ground state candidates corresponding to the embedding (I) can
also be found. They can be described in a way
similar to the discussion for $N=2,3$ to which we
will turn next.

Let us now address the realistic case of $SU(3)_C$. There are two
inequivalent embeddings of $SU(2)_C$ according to which the gluon
octet transforms as
\begin{eqnarray}\label{9}
8\quad&\to&\quad 3+5\qquad\qquad\qquad(I)\nonumber\\
8\quad&\to&\quad 1+2+2+3\qquad(II)\end{eqnarray}
For the first embedding the fundamental three-dimensional
representation (the
quarks) transform as a triplet with respect to the $SO(3)$ subgroup
of $SU(3)$.  The second embedding corresponds  to the
$SU(2)$-``isospin'' symmetry and the results based on this embedding
can be applied to the case $N=2$ as well. In four dimensions there is
obviously no possible choice of a constant gauge field $A^z_\mu$ as
in
(\ref{6}) since the representation (2,2) with respect to $SO(4)$
space rotations cannot be matched with any of the representations
(\ref{9}). A constant field $A_\mu$ does not contain a singlet with
respect to $SU(2)_L\times SU(2)_R$. There is a mismatch between a
large
number of dimensions $(d=4$ in the present case) and a small number
of
colours $(N_C=2,3)$.

The possibility of a constant gauge field exists, nevertheless, for
the three-dimensional theory, which is
relevant as an effective theory at high temperature.
Here the rotation symmetry is reduced to $SO(3)$ and it is now easy
to find a
rotation and translation invariant ground state
\begin{eqnarray}\label{10}
A^\alpha_i&=&a\delta^\alpha_i\nonumber\\
A_0&=&s\end{eqnarray}
with $a,s$ constant. Here $\alpha$ denotes the vector index of an
$SU(2)$-triplet contained in the decomposition (\ref{9}) (or
the vector of an $SU(2)$ gauge group) and  $A_0$ is in the singlet
direction. (We denote by 0 the index corresponding to Euclidean
time and $i$ runs from one to three.) We observe that for $a\not=0$
one can have a non-vanishing $s$ only for the embedding II. (For the
pure three dimensional Yang-Mills theory there is no field $A_0$.)

In four dimensions the state (\ref{10}) is not an acceptable ground
state for the zero temperature theory since  it is not invariant with
respect to ``boosts'' (or the full $SO(4)$ rotations). It is,
however,
a candidate for the ground state at high temperatures
where the $SO(4)$ symmetry is not respected for ``Euclidean time''
compactified on a torus. We observe that the configuration
(\ref{10}) gives rise to non-vanishing constant colour magnetic
fields $\sim a^2$ whereas the colour electric fields vanish for
$s=0$. If there is a transition from a zero temperature ground
state with a symmetry between electric and magnetic fields to the
``asymmetric state'' (\ref{10}) at high temperature the difference
between the electric and magnetic condensate could be an
interesting signal.
\bigskip

3. For the search of a Poincar\'e invariant state for $SU(3)_C$
in four dimensions we have to abandon $A_\mu=$const. We have  to
consider space dependent $A_\mu(x)$ for which  the general
discussion becomes more complicated. Let us first ask under what
conditions we can have  at least a constant field strength
\be\label{11}
F^z_{\mu\nu}=\partial_\mu A^z_\nu-\partial_\nu  A^z_\mu+gf^z_{wy}
A^w_\mu A^y_\nu
\ee
which is invariant under a suitable combination of Lorentz
rotations and global gauge transformations. As we will see the
requirement of an invariant constant field strength is weaker than
the corresponding one  for the gauge field. Invariance of the
ground state field strength under a suitably modified Poincar\'e
transformation is necessary for obtaining Poincar\'e-covariant
Green functions. If such a field strength is found, one may, in a
second step,
attempt the construction of a ground state gauge field which
guarantees the covariance of the Green functions.

With respect to $SU(2)_L\times SU(2)_R$ the field strength
$F^z_{\mu\nu}$ decomposes into two irreducible  representations
$G^z_{\mu\nu}$ and $H^z_{\mu\nu}$ transforming as (3,1) and (1,3):
\begin{eqnarray}\label{12}
G^z_{\mu\nu}&=&\frac{1}{2}(F^z_{\mu\nu}+\tilde
F^z_{\mu\nu})\nonumber\\
H^z_{\mu\nu}&=&\frac{1}{2}(F^z_{\mu\nu}-\tilde F^z_{\mu\nu})
\nonumber\\
\tilde F^z_{\mu\nu}&=&-\frac{1}{2}\varepsilon_{\mu\nu}^{\ \
\rho\sigma}
F^z_{\rho\sigma}\end{eqnarray}
with $\varepsilon_{\mu\nu\rho\sigma}$ the totally antisymmetric
tensor $\varepsilon_{1234}\equiv\varepsilon_{1230}=1$.\footnote{We
use often the index 0 instead of 4 to be close to a Minkowski
notation. Greek indices always run from 1 to 4 (or 0) whereas latin
indices run from 1 to 3.} In terms of colour-magnetic and
colour-electric fields
\begin{eqnarray}\label{13}
F^z_{oi}=E_i^z\qquad &,& \quad \tilde F^z_{oi}=B_i^z\nonumber\\
F^z_{jk}=\varepsilon^{\ \ i}_{jk} B^z_i\ &,&\quad\tilde
F^z_{jk}=\varepsilon_{jk}^{\ \ i} E^z_i
\end{eqnarray}
the self-dual and anti-self-dual fields can be written in the form
\begin{eqnarray}\label{14}
G^z_i&=& G^z_{oi}=\frac{1}{2}\varepsilon_i^{\ jk}
G^z_{jk}=\frac{1}{2}(E_i^z+B_i^z)\nonumber\\
H_i^z&=&H^z_{oi}=-\frac{1}{2}\varepsilon_i^{\ jk}
H^z_{jk}=\frac{1}{2}
(E_i^z-B_i^z).
\end{eqnarray}
Now $G$ is invariant  under $SU(2)_R$ and transforms as a triplet
with respect to $SU(2)_L$. We may therefore form a singlet with
respect to $SU(2)'_L$ if $G^z_{\mu\nu}$ belongs to a triplet
representation $G^\alpha_{\mu\nu}(\alpha=1...3)$ in the
decomposition (\ref{9}). Similar to (\ref{6}) the singlet with
respect to $SU(2)'_L\times SU(2)_R$ reads now
\be\label{15}
B^\alpha_i=E^\alpha_i=b \delta^\alpha_i
\ee
With constant $b$ this configuration  is also translation
invariant. Again, this state with constant magnetic and electric
colour fields is invariant under a modified version of Poincar\'e
symmetry. Since the state (\ref{14}) is self-dual
$(F^z_{\mu\nu}=\tilde F^z_{\mu\nu})$ it follows immediately that it
also obeys the Yang-Mills equation $F^{\mu\nu}_{;\nu}=0$. Written
more
explicitly in terms of the usual $SU(3)_C$ generators (Gell-Mann
matrices $\lambda_z$) the state (\ref{14}) reads for the embeddings
(I), (II) with $F_{\mu\nu}=F^z_{\mu\nu}\lambda_z$
\begin{eqnarray}\label{16}
F_{\mu\nu}&=&-ib\left( \begin{array}{ccc}
0&s_3&-s_2\\
-s_3&0&s_1\\
s_2&-s_1&0\end{array}\right)\qquad (I)\nonumber\\
F_{\mu\nu}&=&b\left( \begin{array}{ccc}
s_3&s_1-is_2&0\\
s_1+is_2&-s_3&0\\
0&0&0\end{array}\right)\qquad (II)\end{eqnarray}
\begin{eqnarray}\label{17}
s_i&=&\delta_{0\mu}\delta_{i\nu}-\delta_{i\mu}\delta_{0\nu}
\nonumber\\
&& +\varepsilon_{ijk}(\delta^j_\mu\delta^k_\nu-\delta^k_\mu
\delta^j_\nu)
\end{eqnarray}
 From the transformation property of the gluon field strength as
(2,2,8) with respect to $SU(2)_L\times SU(2)_R\times SU(3)_C$ we
obtain the representations  of the modified ``Lorentz'' group
$SU(2)'_L\times SU(2)_R$ for the two embeddings (\ref{9})
\begin{eqnarray}\label{18}
&&\begin{array}{lll}
(3,1,8) &\to &(1,1)+(3,1)+(3,1)+(5,1)+(5,1)+(7,1)\\
(1,3,8) &\to &(3,3)+(5,3)\end{array}\Biggr\}\quad (I)\\
&&\begin{array}{lll}
(3,1,8) &\to &(1,1)+(2,1)+(2,1)+(3,1)+(3,1)+(4,1)+(4,1)+(5,1)\\
(1,3,8) &\to &(1,3)+(2,3)+(2,3)+(3,3)\end{array}\Biggr\}
\quad (II)\nonumber\end{eqnarray}
The first striking observation is the appearance  of spin $1/2$ and
spin $3/2$ states with respect to the modified three dimensional
rotation group diag $(SU(2)'_L\times SU(2)_R)$ for the embedding
(II). If we identify the representations (\ref{18}) with (parts of)
the glueball spectrum the embedding (II) would predict not only
bosonic but also fermionic glue ball states! This partial
boson-fermion transmutation would also occur for the embedding (II)
in (\ref{5}). In contrast, the embedding (I) predicts integer spin
for all states in (\ref{18}) (and similar for the
embedding (I) in (\ref{5})). In addition, the $SU(3)_C$ group is
here completely broken without any residual $U(1)_C$ symmetry
commuting with $SU(2)'_L\times SU(2)_R$. With respect to the three
dimensional rotation group diag $(SU(2)'_L\times SU(2)_R)$ the
fields in $G_{\mu\nu}$ and $H_{\mu\nu}$ have the same spectrum,
i.e. singlets, vectors, spin 2 and spin 3 tensors for the
embedding I and additional half integer spin states for the
embedding II. With respect to the modified boosts, however, the
fields in $G_{\mu\nu}$ and $H_{\mu\nu}$ transform quite
differently. This is not a worry of principle since colour neutral
bound states will always have standard Lorentz-transformation
properties. It points, nevertheless, to a strong violation of
left-right symmetry and one wonders how such a spectrum can be
consistent with parity.

4. In addition to the Poincar\'e symmetry we will now also require
 parity conservation for the ground state. Again, we envisage
the possibility that parity is realized as a modified transformation
in combination with a suitable discrete gauge transformation. The
standard parity transformation $(x\to x')$
\begin{eqnarray}\label{19}
x_i\to-x_i,&&\quad\partial_i\to-\partial_i\nonumber\\
x_0\to x_0,&&\quad\partial_0\to \partial_0\end{eqnarray}
reverses the sign of the electric fields
\begin{eqnarray}\label{20}
P(A_i(x))&=&-A_i(x')\nonumber\\
P(A_0(x))&=&A_0(x')\nonumber\\
P(E_i(x))&=&-E_i(x')\nonumber\\
P(B_i(x))&=&B_i(x')\end{eqnarray}
It therefore maps
\begin{eqnarray}\label{21}
G^z_i&\to&-H^z_i\nonumber\\
H^z_i&\to&-G_i^z\end{eqnarray}
and is obviously violated by the state (\ref{15}). Let us ask if
there could be a modified parity transformation with the property
(\ref{19}) and leaving the state (\ref{15}) invariant. (Such a
transformation would belong
to the class $\tilde P$ of generalized parity transformations
discussed in \cite{5}.) The existence of such a transformation
requires a discrete symmetry of the action which does not act
on coordinates but nevertheless implies a mapping $G\to H$.
The modified parity transformation would then be a combination
of this symmetry with the standard parity reflection $P$. Any such
mapping $G\to H$ must act as an automorphism of the group $SO(4)$
exchanging the role of $SU(2)_L$ and $SU(2)_R$. It cannot
be a subgroup of $SO(4)$ since those transformations cannot
``switch''
from one representation to another. The same holds for global
gauge transformations (the latter commute with $SO(4)$.)

In the case of QCD the only symmetry transformation
exchanging $SU(2)_L$ and $SU(2)_R$ representations
also acts as a reflection of an odd number of components of any
$SO(4)$ vector. This holds in particular for the coordinate vector
$x_\mu$ in contradiction to what we are looking for. We conclude
that the state (\ref{15}) spontaneously breaks all possible
generalized
parity symmetries $\tilde P$. It could correspond to the ground
state of a
parity-violating  theory (like a Yang-Mills theory with a chiral
fermion content or explicit parity violating $F_{\mu\nu}\tilde F
^{\mu\nu}$ interactions). It is, however, not a realistic
candidate for QCD which is known to conserve parity. In consequence,
a realistic QCD ground state cannot have a constant field strength
$F_{\mu\nu}$ either!

We should mention at this place that the parity problem is absent for
the
ground state candidate (\ref{6})  for $SU(4)_C$ (or other gauge
groups containing $O(4)$). In fact, the configuration (\ref{6})
violates the standard parity transformation $P$. We may nevertheless
combine $P$ with an automorphism of the $SO(4)$ gauge group which
reverses the sign of three components of $A^\alpha_\mu$, i.e.

$A^\alpha_\mu\to-A^\alpha_\mu$ for $\alpha=1...3$. The state
(\ref{6}) is left invariant by the combined reflection. The global
$SU(4)_C$ gauge transformations contain the required automorphism.
For $SU(4)_C$ one may also construct a ground state candidate
based on the embedding (I) (\ref{5}) by having a nonvanishing
$G_{\mu\nu}$ for the (3,1) representation according to (\ref{15})
and similarly nonvanishing $H_{\mu\nu}$ for the (1,3) representation.
The modified parity reflection is again combined from the standard
parity $P$ and a suitable discrete gauge transformation.
We remember that such a state would have constant
$F_{\mu\nu}$, but not constant $A_\mu$.

5. We may summarize the preceding observations by the statement that
neither
the gauge field nor the field strength can be constant for a
realistic
ground state of four-dimensional QCD with three colours. This implies
that also translation symmetry cannot be realized in the standard
way.
Standard translations have to be combined with suitable gauge
transformations\footnote{An example of such a combination of standard
translations with global abelian gauge transformations is given by
the
symmetry leaving the spin waves of ref. \cite{4} invariant.}. We
require
that any infinitesimal translation of the ground state gauge field
$A^z_\mu(x)$ can be compensated by a corresponding infinitesimal
gauge
transformation with gauge parameter $\theta^z_\mu(x)$
\be\label{22}
-\partial_\mu A^z_\nu(x)=i\theta^w_\mu(x)(T_w)^z_{\ y}A^y_\nu(x)+
\frac{1}{g}\partial_\nu\theta^z_\mu(x)\ee
The combined transformations form the modified translation group with
generators
\be\label{23}
P_\mu=-i\partial_\mu+\theta^w_\mu(x)T_w\ee
(The modified ``momentum operators'' act in a standard way on
tensors, whereas
for gauge fields the inhomogeneous part of the transformation
(\ref{22}) has
to be included.)

Invariance of the ground state gauge field $A^z_\mu(x)$ under
generalized
Lorentz rotations requires similarly
\begin{eqnarray}\label{24}
&&-x_\mu\partial_\nu A^z_\rho+x_\nu\partial_\mu
A^z_\rho+\delta_{\mu\rho}A^z_\nu-\delta_{\nu\rho}A^z_\mu\nonumber\\
&&=i\eta^w_{\mu\nu}(x)(T_w)^z_{\
y}A^y_\rho+\frac{1}{g}\partial_\rho\eta^z_{\mu\nu}(x)\end{eqnarray}
The corresponding generalized (Lorentz) rotation operators are
\begin{eqnarray}\label{25}
M_{\mu\nu}&=&S_{\mu\nu}+L_{\mu\nu}+G_{\mu\nu}\nonumber\\
L_{\mu\nu}&=&-i(x_\mu\partial_\nu-x_\nu\partial_\mu)\nonumber\\
G_{\mu\nu}&=&\eta^w_{\mu\nu}(x)T_w\end{eqnarray}
Here the ``spin operators'' $S_{\mu\nu}$ correspond to rotations
acting on the
vector index of $A_\mu$ and commute with angular momentum
$L_{\mu\nu}$
and the ``gauge part'' $G_{\mu\nu}$
\be\label{26}
[S_{\mu\nu},L_{\rho\sigma}]=0,\quad [S_{\mu\nu},G_{\rho\sigma}]=0\ee
Both $S_{\mu\nu}$ and $L_{\mu\nu}$ obey separately the $SO(4)$
commutation relations
\be\label{27}
[L_{\mu\nu},L_{\rho\sigma}]=i(\delta_{\mu\rho}L_{\nu\sigma}-
\delta_{\mu\sigma}
L_{\nu\rho}-\delta_{\nu\rho}L_{\mu\sigma}+\delta_{\nu\sigma}L_{\mu\rho
})
\ee
(The spin generators $S_{\mu\nu}$ are linear combinations of
$\tau_L,\tau_R$ in
eq. (\ref{1}).)

In consequence, we associate to every element $l_i$ of the standard
Poincar\'e
group an element $g_i$ of the gauge group $G$ such that the
combination
$g_il_i$ leaves $A_\mu$ invariant. This defines a map
\be\label{28}
f:l_i\longrightarrow g_il_i\ee
With respect to the group multiplication the invariance of $A_\mu$
implies for this map
\be\label{29}
l_1l_2\longrightarrow h_{12}g_1l_1g_2l_2\ee
and it is easy to see that $h_{12}$ must be a gauge transformation
which
leaves $A_\mu$ invariant. (We denote by $H$ the subgroup of gauge
transformations
leaving $A_\mu$ invariant ($h_{12}\in H$).) Let us discuss the case
that
$H$ is trivial (only the identity element) or that the $g_i$ can be
chosen
from a subgroup $\tilde G\subset G$ commuting with $H$. Then we can
put $h_{12}
=1$ in eq. (\ref{29}) and the mapping (\ref{28}) is a group
homomorphism.
The elements $g_i$ are uniquely determined in this case by the
invariance
condition $g_il_i(A_\mu)=A_\mu$. The image of $f$ is then isomorphic
to
the Poincar\'e group $P_4$.

By virtue of this isomorphism the generalized translation generators
$P_\mu$ (23) in the different directions must commute
\begin{eqnarray}\label{30}
[P_\mu,P_\nu]&=&-i(\partial_\mu\theta_\nu^z-\partial_\nu\theta_\mu^z)T
_{wz}=0
\nonumber\\
&&\partial_\mu\theta_\nu^z-\partial_\nu\theta_\mu^z=0\end{eqnarray}
Similarly, the modified ``angular momenta'' $M_{\mu\nu}$ must obey
the same $SO(4)$ commutation relations as $L_{\mu\nu}$. In general
$L_{\mu\nu}$
and $G_{\mu\nu}$ do not commute for $\eta$ depending on $x$, and we
obtain similar to (\ref{30}) the consistency condition
\begin{eqnarray}\label{31}
&&\delta_{\mu\rho}\eta^z_{\nu\sigma}-\delta_{\mu\sigma}\eta^z_{\nu\rho
}
-\delta_{\nu\rho}\eta^z_{\mu\sigma}+\delta_{\nu\sigma}\eta
^z_{\mu\rho}-i\eta_{\mu\nu}\eta^x_{\rho\sigma}f^{\ \
z}_{xy}\nonumber\\
&&+x_\mu\partial_\nu\eta^z_{\rho\sigma}-x_\nu\partial_\mu\eta^z_{\rho\
sigma}
-x_\rho\partial_\sigma\eta^z_{\mu\nu}+x_\sigma\partial_\rho\eta^z_{\mu
\nu}=0
\end{eqnarray}
The operators $M_{\mu\nu}$ and $P_\rho$ must generate the generalized
Poincar\'e group.
{}From
\be\label{32}
[M_{\mu\nu},P_\rho]=i\delta_{\mu\rho}P_\nu-i\delta_{\nu\rho}P_\mu\ee
we get the additional consistency relation
\be\label{33}
\delta_\rho\eta^z_{\mu\nu}-x_\mu\partial_\nu\theta^z_\rho+x_\nu
\partial_\mu\theta^z_\rho-\delta_{\mu\rho}\theta^z_\nu+\delta_{
\nu\rho}
\theta^z_\mu+\eta^x_{\mu\nu}\theta^y_\rho f^{\ \ z}_{xy}=0\ee
The problem of finding simultaneous solutions to the conditions
(30), (31), and (33) is equivalent to the problem of finding
embeddings of the
Poincar\'e group $P_4$ into the infinite-dimensional group generated
by local gauge transformations and standard translations and
(Lorentz) rotations. We have presented before a few simple solutions
for
$N\geq 4$, but the general problem is quite difficult to solve. The
problem of
finding simultaneous solutions to eqs. (22) and (24) amounts to the
problem
of finding gauge fields invariant under the modified $P_4$
transformation.
Any solution of eqs. (22), (24) should automatically fulfil eqs.
(30), (31),
and (33) if $h_{12}=1$ in eq. (29) (see above).

It is amazing to see how the ground state problem for QCD in the
framework
of the Euclidean effective action turns into an interesting but
difficult
group-theoretical problem. It is not yet clear to us what is the best
way
for its solution. The observation may be helpful that all gauge
singlets
contracted from the ground state field $A_\mu$ must be invariant
under the
standard Poincar\'e and parity transformations \footnote{One may
check that
the configuration (16) does not obey (34).}, i.e.
\be\label{34}
F^z_{\ \mu\nu}F^{\ \rho\sigma}_z={\rm
const}(\delta^\rho_\mu\delta^\sigma_\nu
-\delta^\sigma_\mu\delta^\rho_\nu)\ee
or

\be\label{35}
F^z_{\ \mu\nu;\rho}F_z^{\ \mu\nu}=0\ee

6. Even though we have not yet been able to solve the ground state
problem
of four-dimensional QCD based on the gauge group $SU(3)_C$, several
lessons
can be learned from our preliminary investigations. We have
demonstrated
for the gauge group $SU(4)_C$ or larger groups that there exist
indeed translation, rotation, and parity-invariant  states with
nonvanishing
field strength. If such a state can be identified as the ground state
and
the form of the effective action for field configurations in the
vicinity
of this state is known, one can derive the spectrum of excitations
from the second functional variation of the effective action
evaluated at the ground
state. This gives the masses of (some of) the glueballs. In this
context
it is interesting to observe that the excitations of the gauge field
have indeed different spins. The field $A_\mu$ can describe a rather
rich
glueball spectrum which is not restricted to spin one states. On the
other
hand the gauge group may be completely broken. (For certain ground
state candidates a residual global gauge symmetry could also
persist.)
The phenomenon of ``gluon condensates'' can be associated in this
language
with the ``spontaneous symmetry-breaking of the gauge symmetry'' by
non-perturbative effects! In this respect there are analogies with
the Higgs phenomenon: The role of the scalar field is now played by
gauge fields
proportional to the ground state field $A_\mu$. This field
corresponds to a
scalar with respect to the modified Poincar\'e transformations. The
ground
state field is equivalent to the vacuum expectation value of this
``scalar''
excitation\cite{7}. The (generalized) scalar excitation corresponding
to the Higgs
boson is always present within the spectrum of glueball states since
the
gauge field must contain a singlet with respect to the generalized
(Lorentz) rotations.

We have seen that the ``embedding problem'' of finding a generalized
$P_4$
subgroup and a corresponding state left invariant by this subgroup
changes
qualitatively for a small number of colours $N$. If this is connected
to an
important quantitative change in the ground state properties for
$N\geq 4$ and $N<4$ an expansion in $1/N$ for a large number of
colours may sometimes
produce misleading results for $SU(3)_C$.

Another observation is the appearance of more than one ground state
candidate
for $SU(4)_C$. This suggests that the ground state may not be fixed
uniquely
by the requirement of Poincar\'e and parity invariance and
$F_{\mu\nu}\not=0$. In order to distinguish between different
possible candidates and to select
the true ground state one needs details of the effective action. (The
ground
state corresponds to the absolute minimum of the effective action.) A
reliable computation of the effective action is not easy because of
the severe infrared
problems in perturbative QCD. In this context the concept of the
scale-dependent
effective average action $\Gamma_k$ \cite{8} may prove a useful tool.
For the
effective average action only the quantum fluctuations with momenta
larger
than an infrared cutoff, $q^2>k^2$, are included. The effective
action obtains
then in the limit $k\to 0$. The dependence of $\Gamma_k$ on the scale
$k$
is governed by an exact evolution equation \cite{9}. It is
encouraging
to observe in this context that the lowest order invariant
\be\label{36}
\Gamma_k^{(0)}=\frac{Z_{F,k}}{4}\int d^4xF^z_{\ \mu\nu}F_z^{\ \mu\nu}
\ee
changes its sign for small $k$ \cite{10}, i.e.
\be\label{37}
Z_{F,k}<0\quad{\rm for}\quad k<\Lambda_{conf}\ee
Here the vanishing of $Z_{F,k}$ is directly related to a diverging
renormalized gauge coupling and occurs at the confinement scale
$\Lambda_{conf}$. The negative value of $Z_{F,k}$ for small $k$
is a clear sign of the instability of the perturbative vacuum and
the onset of
``gluon condensation'' or ``nonperturbative spontaneous
symmetry-breaking
of the gauge group''. A calculation of the $k$-dependence of the
coefficients
of higher invariants (e.g. $\sim (F^{\ z}_{\mu\nu}F^{\mu\nu}_z)^2$)
is in
progress \cite{11} and should shed some light on the properties of
the ground
state in terms of the values of various gauge invariants formed from
the
ground state field $A_\mu$. In particular, it would be interesting to
know if
a convariantly constant field strength
\be\label{38}
F_{\mu\nu;\rho}=0\ee
is favoured for the ground state or not. Needless to say that a
determination
of the QCD ground state either by group-theoretical methods or by the
use of detailed properties of the effective action would offer new
insights
into various phenomena of the theory of strong interactions.

  \end{document}